\documentclass[reqno ,12pt]{amsart}
\usepackage[foot]{amsaddr}
\usepackage{graphicx}
\usepackage{enumitem}
\usepackage[usenames,dvipsnames]{xcolor}
\usepackage[paperwidth=8.5in,paperheight=11in,text={6in,9in},left=1.25in,top=1in,headheight=0.25in,headsep=0.4in,footskip=0.4in]{geometry}
\usepackage{subfigure}
\usepackage{lineno}
\usepackage{setspace}
\usepackage[longnamesfirst]{natbib}

\usepackage{mathpazo}
\definecolor{linenocolor}{gray}{0.6}
\definecolor{prec}{RGB}{62,135,205}
\definecolor{sens}{RGB}{255,122,0}
\definecolor{spec}{gray}{0.4}
\definecolor{red}{RGB}{255,0,0}

\usepackage{fix-cm}

\usepackage{hyperref}

\makeatletter
\newenvironment{startquote}[2][2em]
  {\setlength{\@tempdima}{#1}%
   \def\chapquote@author{#2}%
   \parshape 1 \@tempdima \dimexpr\textwidth-2\@tempdima\relax%
   \itshape}
  {\par\normalfont\hfill--\ \chapquote@author\hspace*{\@tempdima}\par\bigskip}
\makeatother

\begin{document}

\title[The Natural Selection of Bad Science]{\large The Natural Selection of Bad Science}
\author[Smaldino \& McElreath]{Paul E. Smaldino$^1$ \and Richard McElreath$^{2,3}$}
\address{$^1$Department of Political Science, University of California-Davis, Davis, CA 95616}
\address{$^2$Department of Anthropology, University of California-Davis, Davis,CA 95616}
\address{$^3$Max Planck Institute for Evolutionary Anthropology, Leipzig, Germany}
\email{paul.smaldino@gmail.com}




\maketitle

{\vspace{-6pt}\footnotesize\begin{center}\today\end{center}\vspace{24pt}}

\noindent\textsc{Abstract.} 
Poor research design and data analysis encourage false-positive findings. Such poor methods persist despite perennial calls for improvement, suggesting that they result from something more than just misunderstanding. The persistence of poor methods results partly from incentives that favor them, leading to the natural selection of bad science. This dynamic requires no conscious strategizing---no deliberate cheating nor loafing---by scientists, only that publication is a principle factor for career advancement. Some normative methods of analysis have almost certainly been selected to further publication instead of discovery. 
In order to improve the culture of science, a shift must be made away from correcting misunderstandings and towards rewarding understanding. 
We support this argument with empirical evidence and computational modeling. We first present a 60-year meta-analysis of statistical power in the behavioral sciences and show that power has not improved despite repeated demonstrations of the necessity of increasing power. 
To demonstrate the logical consequences of structural incentives, 
we then present a dynamic model of scientific communities in which competing laboratories investigate novel or previously published hypotheses using culturally transmitted research methods. As in the real world, successful labs produce more ``progeny,'' such that their methods are more often copied and their students are more likely to start labs of their own. Selection for high output leads to poorer methods and increasingly high false discovery rates. We additionally show that replication slows but does not stop the process of methodological deterioration. 
Improving the quality of research requires change at the institutional level.

\vspace{12pt}


\vspace{24pt}
\newpage

\begin{startquote}{Donald T. Campbell (1976, p. 49)}
\noindent The more any quantitative social indicator is used for social decision-making, the more subject it will be to corruption pressures and the more apt it will be to distort and corrupt the social processes it is intended to monitor. \nocite{Campbell:1976}
\end{startquote}

\begin{startquote}{Terry McGlynn (realscientists) (21  October 2015, 4:12 p.m. Tweet.)}
\noindent I've been on a number of search committees. I don't remember anybody looking at anybody's papers. Number and IF [impact factor] of pubs are what counts. 
\end{startquote}

\vspace{-12pt}

\doublespacing 
\section{Introduction}

In March 2016, the American Statistical Association published a set of corrective guidelines about the use and misuse of $p$-values \citep{ASApvalue}. Statisticians have been publishing guidelines of this kind for decades \citep{meehl1967, cohen1994}. Beyond mere significance testing, research design in general has a history of shortcomings and repeated corrective guidelines. Yet misuse of statistical procedures and poor methods has persisted and possibly grown. 
In fields such as psychology, neuroscience, and medicine, practices that increase false discoveries remain not only common, but normative  \citep{enserink2012stapel,vul2009voodoo,peterson2016baby,kerr1998harking,john2012questionable,ioannidis2014toogood,simmons2011false}. 

Why have attempts to correct such errors so far failed? In April 2015, members of the United Kingdom's science establishment attended a closed-door symposium on the reliability of biomedical research \citep{Horton2015}. The symposium focused on the contemporary crisis of faith in research. Many prominent researchers believe that as much as half of the scientific literature---not only in medicine, by also in psychology and other fields---may be wrong \citep{ioannidis_why_2005,simmons2011false,pashler_replicability_2012, mcelreath_boffland_2015}. Fatal errors and retractions, especially of prominent publications, are increasing \citep{ioannidis2005contradicted,ioannidis2012selfcorr,steen2013retractions}. The report that emerged from this symposium echos the slogan of one anonymous attendee: ``Poor methods get results.'' Persistent problems with scientific conduct have more to do with incentives than with pure misunderstandings. So fixing them has more to do with removing incentives that reward poor research methods than with issuing more guidelines. As Richard Horton, editor of \emph{The Lancet}, put it: ``Part of the problem is that no one is incentivised to be right'' \citep{Horton2015}.

This paper argues that some of the most powerful incentives in contemporary science actively encourage, reward, and propagate poor research methods and abuse of statistical procedures. We term this process \emph{the natural selection of bad science} to indicate that it requires no conscious strategizing nor cheating on the part of researchers. Instead it arises from the positive selection of methods and habits that lead to publication. How can natural selection operate on research methodology? There are no research ``genes.'' But science is a cultural activity, and such activities change through evolutionary processes \citep{campbell1965variation, skinner1981selection, boyd1985culture, mesoudi2011cultural, whiten2011culture, smaldino2014cultural, acerbi2015if}. Karl Popper (1979), \nocite{popper1979objective} following Campbell (1965), \nocite{campbell1965variation} discussed how scientific theories evolve by variation and selection retention. But scientific {\em methods} also develop in this way. Laboratory methods can propagate either directly, through the production of graduate students who go on to start their own labs, or indirectly, through prestige-biased adoption by 
researchers in other labs. Methods which are associated with greater success in academic careers will, other things being equal, tend to spread.

The requirements for natural selection to produce design are easy to satisfy. Darwin outlined the logic of natural selection as requiring three conditions:
\begin{enumerate}[noitemsep,topsep=0pt]
\item There must be variation.
\item That variation must have consequences for survival or reproduction.
\item Variation must be heritable.
\end{enumerate}
In this case, there are no biological traits being passed from scientific mentors to apprentices. However, research practices do vary. That variation has consequences---habits that lead to publication lead to obtaining highly competitive research positions. And variation in practice is partly heritable, in the sense that apprentices acquire research habits and statistical procedures from mentors and peers. Researchers also acquire research practice from successful role models in their fields, even if they do not personally know them. Therefore, when researchers are rewarded primarily for publishing, then habits which promote publication are naturally selected. Unfortunately, such habits can directly undermine scientific progress.

This is not a new argument. But we attempt to substantially strengthen it. We support the argument both empirically and analytically. We first review evidence that institutional incentives are likely to increase the rate of false discoveries. Then we present evidence from a literature review of repeated calls for improved methodology, focusing on the commonplace and easily understood issue of statistical power. We show that despite over 50 years of reviews of low statistical power and its consequences, there has been no detectable increase. 

While the empirical evidence is persuasive, it is not conclusive. It is equally important to demonstrate that our argument is logically sound. Therefore we also analyze a formal model of our argument. Inspecting the logic of the selection-for-bad-science argument serves two purposes. First, if the argument cannot be made to work in theory, then it cannot be the correct explanation, whatever the status of the evidence. Second, formalizing the argument produces additional clarity and the opportunity to analyze and engineer interventions. To represent the argument, we define a dynamical model of research behavior in a population of competing agents. We assume that all agents have the utmost integrity. They never cheat. Instead, research methodology varies and evolves due to its consequences on hiring and retention, primarily through successful publication. As a result our argument applies even when researchers do not directly respond to incentives for poor methods. We show that the persistence of poor research practice can be explained as the result of the natural selection of bad science.

\section{Institutional incentives for scientific researchers}

The rate at which new papers are added to the scientific literature has steadily  increased in recent decades. This is partly due to more opportunities for collaboration, resulting in more multi-author papers \citep{nabout2015publish, wardil2015coauthor}. However, the increases in publication rate may also be driven by changing incentives. Recently, Brischoux and Angelier (2015) \nocite{brischoux2015academia} looked at the  career statistics of junior researchers hired by the French CNRS in evolutionary biology between 2005 and 2013. They found persistent increases in the average number of publications at the time of hiring: newly hired biologists now have almost twice as many publications as they did ten years ago (22 in 2013 vs. 12.5 in 2005).  
These numbers reflect intense competition for academic research positions. The world's universities produce many more PhDs than there are permanent academic positions for them to fill \citep{cyranoski2011education, schillebeeckx2013missing, powell2015future}, and while this problem has escalated in recent years, it has been present for at least two decades \citep{kerr1995too}. Such competition is all the more challenging for researchers who graduate from any but the most prestigious universities, who face additional discrimination on the job market \citep{clauset2015systematic}. Although there may be jobs available outside of academia---indeed, often better-paying jobs than university professorships---tenure-track faculty positions at major research universities come with considerable prestige, flexibility, and creative freedom, and remain desirable. Among those who manage to get hired, there is continued competition for grants, promotions, prestige, and placement of graduate students. 

Given this competition, there are incentives for scientists to stand out among their peers. Only the top graduate students can become tenure-track professors, and only the top assistant professors will receive tenure and high profile grants. Recently, the Nobel laureate physicist Peter Higgs, who pioneered theoretical work in the search for fundamental particles in the 1960s, lamented ``Today I wouldn't get an academic job. It's as simple as that. I don't think I would be regarded as productive enough" (The Guardian, 6 Dec 2013). Of course, such a statement is speculative; a young Peter Higgs might well get a job today. But he might not be particularly successful. Evidence suggests that only a very small proportion of scientists produce the bulk of published research and generate the lion's share of citation impact \citep{ioannidis2014estimates}, and it is these researchers who are likely the source of new labs and PhDs. Supporting this argument is evidence that most new tenure-track positions are filled by graduates from a small number of elite universities---typically those with very high publication rates \citep{clauset2015systematic}. 

One method of distinguishing oneself might be to portray one's work as groundbreaking. And indeed, it appears that the innovation rate has been skyrocketing. Or claims at innovation, at any rate. In the years between 1974 and 2014, the frequency of the words ``innovative," "groundbreaking," and ``novel" in PubMed abstracts increased by 2500\% or more \citep{vinkers2015}. As it is unlikely that individual scientists have really become 25 times more innovative in the past 40 years, one can only conclude that this language evolution reflects a response to increasing pressures for novelty, and more generally to stand out from the crowd. 

Another way to distinguish oneself is through sheer volume of output. Substantial anecdotal evidence suggests that number of publications is an overwhelmingly important factor in search committee decision making. Output may be combined with impact---some researchers place emphasis on metrics such the $h$-index, defined as the largest number $h$ such that an individual has $h$ publications with at least $h$ citations each \citep{hirsch2005index}. Yet volume alone is often perceived as a measure of researcher quality, particularly for early-career researchers who have not yet had the time to accrue many citations. Although the degree to which publication output is used for evaluation---as well as what, exactly, constitutes acceptable productivity---varies by discipline, our argument applies to all fields in which the number of published papers, however scaled, is used as a benchmark of success. 

Whenever a quantitative metric is used as a proxy to assess a social behavior, it becomes open to exploitation and corruption \citep{Campbell:1976, goodhart1975problems, lucas1976econometric}. This is often summarized more pithily as ``when a measure becomes a target, it ceases to be a good measure." For example, since the adoption of the $h$-index, researchers have been observed to artificially inflate their indices through self-citation \citep{bartneck2011detecting} and even a clever type of quasi-fraud. With the goal of illustrating how the $h$-index system might be gamed, researchers created six fake papers under a fake author that cited their papers extensively \citep{delgado2014google}. They posted these papers to a university server. When the papers were indexed by Google, their $h$-indices on Google Scholar increased dramatically. 
Strategies that target evaluative metrics 
may be invented by cheaters, but they may propagate through their consequences\footnote{Incentives to increase one's $h$-index may also encourage researchers to engage in high-risk hypothesizing, particularly on ``hot" research topics, because they can increase their citation count by being corrected.}.

Our argument is that incentives to generate a lengthy CV have particular consequences on the ecology of scientific communities. The problem stems, in part, from the fact that positive results in support of some novel hypothesis are more likely to be published than negative results, particularly in high-impact journals.  For example, until recently, the {\em Journal of Personality and Social Psychology} refused to publish failed replications of novel studies it had previously published \citep{aldhous2011Bem}. Many researchers who fail to garner support for their hypotheses do not even bother to submit their results for publication \citep{franco_publication_2014}. The response to these incentives for positive results is likely to increase false discoveries. 

If researchers are rewarded for publications and positive results are generally both easier to publish and more prestigious than negative results, then researchers who can obtain more positive results---whatever their truth value---will have an advantage. Indeed, researchers sometimes fail to report those hypotheses that fail to generate positive results (lest reporting such failures hinder publication) \citep{franco_publication_2015}, even though such practices make the publication of false positives more likely \citep{simmons2011false}. One way to better ensure that a positive result corresponds to a true effect is to make sure one’s hypotheses have firm theoretical grounding and that one’s experimental design is sufficiently well powered \citep{mcelreath_boffland_2015}. However, this route takes effort, and is likely to slow down the rate of production. An alternative way to obtain positive results is to employ techniques, purposefully or not, that drive up the rate of false positives. 
Such methods have the dual advantage of generating output at higher rates than more rigorous work, while simultaneously being more likely to generate publishable results. Although sometimes replication efforts can reveal poorly designed studies and irreproducible results, this is more the exception than the rule \citep{schmidt_shall_2009}. For example, it has been estimated that less than 1\% of all psychological research is ever replicated \citep{makel_replications_2012}, and failed replications are often disputed \citep{bissell_reproducibility_2013, schnall_clean_2014, kahneman_new_2014}. Moreover, even firmly discredited research is often cited by scholars unaware of the discreditation \citep{campanario2000fraud}. Thus, once a false discovery is published, it can permanently contribute to the metrics used to assess the researchers who produced it.   

False discoveries in the literature can obviously result from fraud or $p$-hacking \citep{aschwanden2015science}, but there are many ways that false discoveries can be generated by perfectly well-intentioned researchers. These are easy to spot when the results are absurd; for example, following standard methods for their fields, researchers have observed a dead Atlantic Salmon exhibiting neural responses to emotional stimuli \citep{bennett2009neural} and university students apparently demonstrating ``pre-cognitive" abilities to predict the outcome of a random number generator \citep{wagenmakers2011psychologists}. However, false discoveries are usually not identifiable at a glance, which is why they are problematic. In some cases, poor or absent theory amounts to hypotheses being generated almost at random, which substantially lowers the probability that a positive result represents a real effect \citep{ioannidis_why_2005, mcelreath_boffland_2015, gigerenzer1998surrogates}. Interpretation of results after data is collected can also generate false positives through biased selection of statistical analyses \citep{gelman2014statistical} and post-hoc hypothesis formation \citep{kerr1998harking}. 

Campbell’s Law, stated in this paper's epigraph, implies that if researchers are incentivized to increase the number of papers published, they will modify their methods to produce the largest possible number of publishable results rather than the most rigorous investigations. We propose that this incentivization can create selection pressures for the cultural evolution of poor methods that produce publishable findings at the expense of rigor. 
It is important to recognize that this process does not require any direct strategizing on the part of individuals. To draw an analogy from biological evolution, giraffe necks increased over time, not because individual animals stretched their necks, but because those with longer necks could more effectively monopolize food resources and thereby produce more offspring. In the same way, common methodologies in scientific communities can change over time not only because established researchers are strategically changing their methods, but also because certain researchers are more successful in transmitting their methods to younger generations. Incentives influence both the patterns of innovation and the nature of selection. Importantly, it is not necessary that strategic innovation be common in order for strategic innovations to dominate in the research population.

\section{Case study: Statistical power has not improved}

As a case study, let us consider the need for increased statistical power. Statistical power refers to the probability that a statistical test will correctly reject the null hypothesis when it is false, given information about sample size, effect size, and likely rates of false positives\footnote{We differentiate {\em statistical power} from power more generally; the latter is the probability that one's methods will return a positive result given a true effect, and is a Gestalt property of one's methods, not only of one's statistical tools.} \citep{cohen1992statistical}. Because many effects in the biomedical, behavioral, and social sciences are small \citep{richard2003quality, aguinis2005effect, kuhberger2014publication}, it is important for studies to be sufficiently high-powered. On the other hand, low-powered experiments are substantially easier to perform when studying human or other mammals, particularly in cases where the total subject pool is small, the experiment requires expensive equipment, or data must be collected longitudinally. It is clear that low-powered studies are more likely to generate false negatives. Less clear, perhaps, is that low power can also increase the false discovery rate and the likelihood that reported effect sizes are inflated, due to their reduced ability to mute stochastic noise \citep{ioannidis_why_2005, button_power_2013,LakensEvers:2014}. 

The nature of statistical power sets up a contrast. In an imaginary academic environment with purely cooperative incentives to reveal true causal models of nature, increasing power is often a good idea. Infinite power makes no sense. But very low power brings no benefits. However, in a more realistic academic environment that only publishes positive findings and rewards publication, an efficient way to succeed is to conduct low power studies. Why? Such studies are cheap and can be farmed for significant results, especially when hypotheses only predict differences from the null, rather than precise quantitative differences and trends \citep{meehl1967}. We support this prediction in more detail with the model in a later section---it is possible for researchers to publish more by running low-power studies, but at the cost of filling the scientific literature with false discoveries. For the moment, we assert the common intuition that there is a conflict of interest between the population and the individual scientist over statistical power. The population benefits from high power more than individual scientists do. Science does not possess an ``invisible hand'' mechanism through which the naked self-interest of individuals necessarily brings about a collectively optimal result.

Scientists have long recognized this conflict of interest. The first highly-cited exhortation to increase statistical power was published by Cohen in 1962, as a reaction to the alarmingly low power of most psychology studies at the time \citep{cohen1962statistical}. The response, or lack of response, to such highly-cited exhortations serves as a first-order test of which side of the conflict of interest is winning. A little over two decades after Cohen's original paper, two meta-analyses by Sedlmeier and Gigerenzer (1989) \nocite{sedlmeier1989} and Rossi (1990) \nocite{rossi1990statistical} examined a total of 25 reviews of statistical power in the psychological and social science literature between 1960 and 1984. These studies found that not only was statistical power quite low, but that in the intervening years since Cohen (1962), \nocite{cohen1962statistical} no improvement could be discerned. Recently, 
\cite{vankov2014} observed that statistical power in psychological science appears to have remained low to the present day.

\begin{figure}[tp]
\begin{center}
	\hspace{-0.5in}\includegraphics[width=4in]{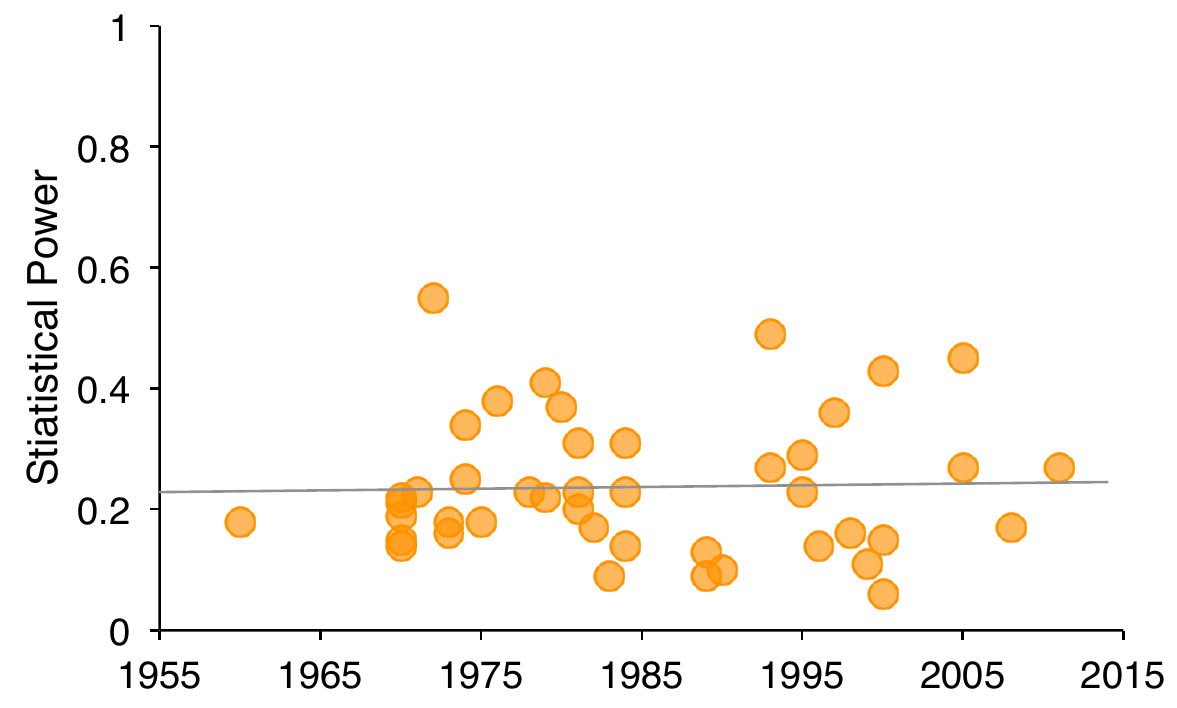}
\caption{Average statistical power from 44 reviews of papers published in journals in the social and behavioral sciences between 1960 and 2011. Data are power to detect small effect sizes ($d = 0.2$), assuming a false positive rate of $\alpha = 0.05$, and indicate both very low power (mean = 0.24) but also no increase over time ($R^2 = 0.00097$).}
\label{fig_powerdata}
\end{center}
\end{figure}

We expanded this analysis by performing a search on Google Scholar among papers that had cited Sedlmeier and Gigerenzer (1989) \nocite{sedlmeier1989} (the more highly cited of the two previous meta-analyses) using the search terms ``statistical power" and ``review." We collected all papers that contained reviews of statistical power from published papers in the social, behavioral, and biological sciences, and found 19 studies from 16 papers published between 1992 and 2014. Details about our methods and data sources can be found in the Appendix. We focus on the statistical power to detect small effects of the order $d = 0.2$, the kind most commonly found in social science research. These data, along with the data from Sedlmeier and Gigerenzer (1989) \nocite{sedlmeier1989} and Rossi (1990) \nocite{rossi1990statistical}, are plotted in Figure \ref{fig_powerdata}. Statistical power is quite low, with a mean of only 0.24, meaning that tests will fail to detect small effects when present three times out of four. More importantly, statistical power shows no sign of increase over six decades ($R^2 = 0.00097$). The data are far from a complete picture of any given field or of the social and behavioral sciences more generally, but they help explain why false discoveries appear to be common. Indeed, our methods may {\em overestimate} statistical power because we draw only on published results, which were by necessity sufficiently powered to pass through peer review, usually by detecting a non-null effect\footnote{It is possible that the average statistical power of research studies does, in fact, sometimes increase for a period, but is then quelled as a result of publication bias. Publication in many disciplines is overwhelmingly biased toward non-null results. Therefore, average power could increase (at least in the short run), but, especially if hypothesis selection did not improved at a corresponding rate, it might simply lead to the scenario in which higher-powered studies merely generated more null results, which are less likely to be published \citep{nosek2012scientific}. Labs employing lower-powered studies would therefore have an advantage, and lower-powered methods would continue to propagate.}.

Why does low power, a conspicuous and widely-appreciated case of poor research design, persist? There are two classes of explanations. First, researchers may respond directly to incentives and strategically reason that these poor methods help them maximize career success. In considering the persistence of low statistical power, 
\cite{vankov2014} suggest the following: ``Scientists are human and will therefore respond (consciously or unconsciously) to incentives; when personal success (e.g., promotion) is associated with the quality and (critically) the {\em quantity} of publications produced, it makes more sense to use finite resources to generate as many publications as possible” (p. 1037,  emphasis in original). Second, researchers may be trying to do their best, but selection processes reward misunderstandings and poor methods. Traditions in which people believe these methods achieve broader scientific goals will have a competitive edge, by crowding out alternative traditions in the job market and limited journal space. 
Vankov et al. provide some evidence for this among psychologists: Widespread misunderstandings of power and other statistical issues. What these misunderstandings have in common is that they all seem to share the design feature of making positive results---true or false---more likely. Misunderstandings that hurt careers are much less commonplace.

Reality is probably a mix of these explanations, with some individuals and groups exhibiting more of one than the other. Our working assumption is that most researchers have internalized scientific norms of honest conduct and are trying their best to reveal true explanations of important phenomena. However, the evidence available is really insufficient. Analyses of data in evolutionary and historical investigations are limited in their ability infer dynamical processes \citep{smaldino_abm_2015}, particularly when those data are sparse, as with investigations of scientific practices. To really investigate such a population dynamic hypothesis, we need a more rigorous demonstration of its logic.

\section{An Evolutionary Model of Science}

To validate the logic of the natural selection of bad science, we develop and analyze a  dynamical population model. Such a model simultaneously verifies the logic of a hypothesis and helps to refine its predictions, so that it can be more easily falsified \citep{wimsatt1987falsemodels,epstein2008whymodel}. Our model is evolutionary: researchers compete for prestige and jobs in which the currency of fitness is number of publications, and more successful labs will have more progeny that inherit their method. 

The foundation is a previously published mathematical model 
\citep{mcelreath_boffland_2015} in which a population of scientists investigate both novel and previously tested hypotheses and attempt to communicate their results to produce a body of literature. Variation in research quality, replication rates, and publication biases are all present in the dynamics. That model was defined analytically and solved exactly for the probability that a given hypothesis is true, conditional on any observed publication record.

Here we extend this model to focus on a finite, heterogeneous population of $N$ labs. We assume the following: 
\begin{itemize}
\item Each lab has a characteristic {\em power}, the ability to positively identify a true association. This power is not only the formal power of a statistical procedure. Rather it arises from the entire chain of inference.
\item Increasing power also increases the rate of false positives, unless {\em effort} is exerted. 
\item Increasing effort decreases the productivity of a lab, because it takes longer to perform rigorous research. 
\end{itemize}
It is important to understand why increasing power tends to also increase false positives without the application of effort. It is quite easy to produce a method with very high power: simply declare support for every proposed association, and you are guaranteed to never mistakenly mark a true hypothesis as false. Of course, this method will also yield many false positives. One can decrease the rate of false positives by requiring stronger evidence to posit the existence of an effect. However, doing so will also decrease the power---because even true effects will sometimes generate weak or noisy signals---unless effort is exerted to increase the size and quality of one’s dataset. This follows readily from the logic of signal detection theory \citep{macmillan1991detection}. 

There are alternative ways to formalize this trade-off. For example, we might instead assume signal threshold and signal noise to be characteristics of a lab. That would hew closer to a signal detection model. What we have done here instead is focus on a behavioral hypothesis, that researchers tend to reason \emph{as if} a research hypothesis were true and select methods that make it easier to find true effects. This maintains or increases effective power, even if nominal statistical power is low. But it simultaneously exaggerates false-positive rates, because many hypotheses are in fact not true. Formally, either system could be translated to the other, but each represents a different hypothesis about the behavioral dimensions along which inferential methods change. The trade-off between effective power and research effort has been invoked in less formal arguments, as well \citep[p. 288]{LakensEvers:2014}. 

Given their inferential characteristics, labs perform experiments and attempt to publish their results. But positive results are easier to publish than negative results. Publications are rewarded (through prestige, grant funding, and more opportunities for graduate students), and more productive labs are in turn more likely to propagate their methodologies onto new labs (such as those founded by their successful grad students). New labs resemble, but are not identical to, their parent labs.

The model has two main stages: {\em Science} and {\em Evolution}. In the Science stage, each lab has the opportunity to select a hypothesis, investigate it experimentally, and attempt to communicate their results through a peer-reviewed publication. Hypotheses are assumed to be strictly true or false, though their essential epistemological states cannot be known with certainty but can only be estimated using experiments. In the Evolution stage, an existing lab may ``die" (cease to produce new research), making room in the population for a new lab that adopts the methods of a progenitor lab. More successful labs are more likely to produce progeny. 

\subsection{Science}
The Science stage consists of three phases: hypothesis selection, investigation, and communication. 
Every time step, each lab $i$, in random order, begins a new investigation with probability $h(e_i)$, where $e_i$ is the characteristic {\em effort} that the lab exerts toward using high quality experimental and statistical methods.  Higher effort results in better methods (more specifically, it allows higher power for a given false positive rate), but also results in a lengthier process of performing and analyzing experiments\footnote{We acknowledge that even methods with low power and/or high rates of false positives may require considerable time and energy to apply, and might therefore be considered effortful in their own right. For readers with such an objection, we propose substituting the word {\em rigor} to define the variable $e_i$.}. 
The probability of tackling a new hypothesis on a given time step is: 
\begin{equation}
h(e_i) = 1 - \eta \log_{10} e_i 
\end{equation}
where $\eta$ is a constant reflecting the extent to which increased effort lowers the lab's rate of producing new research. For simplicity, $e_i$ is bounded between 1 and 100 for all labs. In all our simulations $\eta = 0.2$, which ensured that $h$ stayed non-negative. This value is fairly conservative; in most of our simulations, labs were initialized with a fairly high rate of investigation, $h = 0.63$, at their highest level of effort. However, our results are robust for any monotonically decreasing, non-negative function $h(e_i)$. 

If an experimental investigation is undertaken, the lab selects a hypothesis to investigate. With probability $r_i$, this hypothesis will be one that has been supported at least once in the published literature---i.e., it will be an attempt to replicate prior research. Otherwise, the lab selects a novel, untested hypothesis (at least as represented in the literature) for investigation. Novel hypotheses are true with probability $b$, the base rate for any particular field. It is currently impossible to accurately calculate the base rate; it may be as high as 0.1 for some fields but it is likely to be much lower in many others \citep{ioannidis_why_2005, pashler_replicability_2012, mcelreath_boffland_2015}. 

\begin{figure}[tp]
\begin{center}
	\hspace{-0.5in}\includegraphics[width=4in]{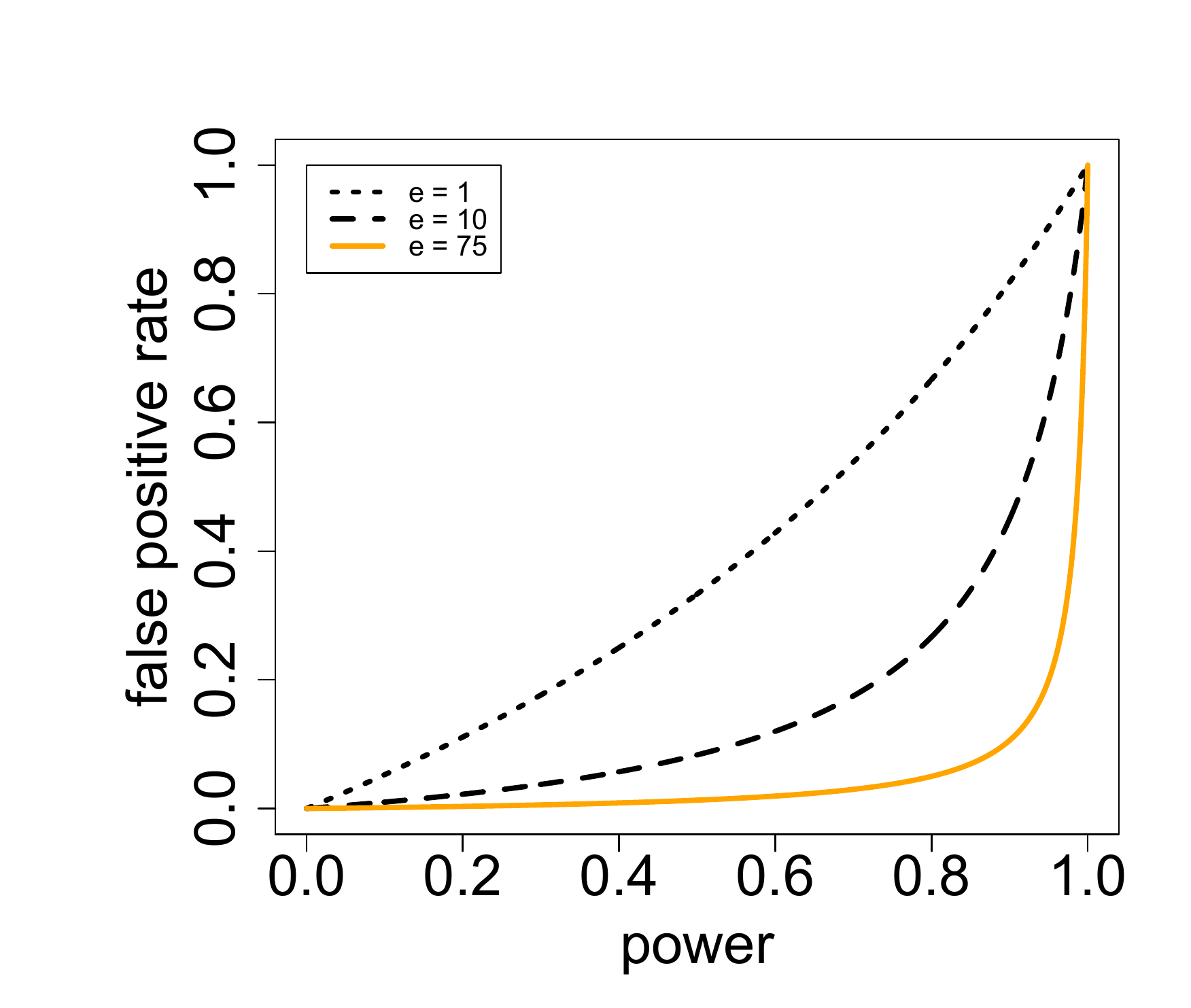}
\caption{The relationship between power and false positive rate, modified by effort, $e$. Runs analyzed in this paper were initialized with $e_0$ = 75 (shown in orange), such that $\alpha = 0.05$ when power is 0.8.}
\label{fig_signaldetection}
\end{center}
\end{figure}

Labs vary in their investigatory approach. Each lab $i$ has a characteristic power, $W_i$ associated with its methodology, which defines its probability of correctly detecting a true hypothesis, $\Pr(+|T)$. Note again that power here is a characteristic of the entire investigatory process, not just the statistical procedure. It can be very high, even when sample size is low. 
The false positive rate, $\alpha_i$, must be a convex function of power. We assume it to be a function of both power and the associated rigor and effort of the lab's methodology: 
\begin{equation}
\alpha_i = \frac{W_i}{1 + (1 - W_i) e_i}.
\end{equation}
This relationship is depicted in Figure~\ref{fig_signaldetection}. What this functional relationship reflects is the necessary signal-detection trade off: finding all the true hypotheses necessitates labeling all hypotheses as true. Likewise, in order to never label a false hypothesis as true, one must label all hypotheses as false. Note that the {\em false discovery rate}---the proportion of positive results that are in fact false positives---is determined not only by the false positive rate, but also by the base rate, $b$. When true hypotheses are rarer, false discoveries will occur more frequently \citep{ioannidis_why_2005,mcelreath_boffland_2015}.

A feature of our functional approach is that increases in effort do not also increase power; the two variables are independent in our model. This is unexpected from a pure signal detection perspective. Reducing signal noise, by increasing experimental rigor, will tend to influence both true and false positive rates. However, our approach makes sense when power is maintained by contingent procedures that are invoked conditional on the nature of the evidence \citep{gelman2014statistical}. 
If we instead view researchers' inferential procedures as fixed, prior to seeing the data, the independence of effort and power is best seen as a narrative convenience: effort is the sum of all methodological behaviors that allow researchers to increase their power without also increasing their rate of false positives.

All investigations yield either positive or negative results: a true hypothesis yields a positive result with probability $W_i$, and a false hypothesis yields a positive result with probability $\alpha_i$. Upon obtaining these results, the lab attempts to communicate them to a journal for publication. We assume that positive novel results are always publishable, while negative novel results never are. Both confirmatory and disconfirmatory replications are published, possibly at lower rates. We adopt this framework in part because it approximates widespread publication practices. Positive and negative replications are communicated with probabilities $c_{R+}$ and $c_{R-}$, respectively. 

Communicated results enter the published literature. Labs receive payoffs for publishing their results, and these payoffs---which may be thought of in terms of factors such as prestige, influence, or funding---make their methodologies more likely to propagate in the scientific community at large. Labs accumulate payoffs throughout their lifespans. Payoffs differ for novel and replicated results, with the former being larger.  Payoffs can also accrue when other research groups attempt to replicate a lab's original hypothesis. These payoffs can be positive, in cases of confirmatory replication, or punitive, in cases of failure to replicate.  Values for payoffs and all other parameter values are given in Table \ref{table_parameters}. 

\subsection{Evolution} 
At the end of each time step, once all labs have had the opportunity to perform and communicate research, there follows a stage of selection and replication. First, a lab is chosen to die. A random sample of $d$ labs is obtained, and the oldest lab of these is selected to die, so that age correlates coarsely but not perfectly with fragility.  If multiple labs in the sample are equally old, one of these is selected at random. The dying lab is then removed from the population. Next, a lab is chosen to reproduce. A random sample of $d$ labs is obtained, and from among these the lab with the highest accumulated payoff is chosen to reproduce. This skews reproduction toward older labs as well as toward more successful labs, which agrees with the observation that more established labs and scientists are more influential. However, because age does not correlate with the {\em rate} at which payoffs are accumulated, selection will favor those strategies which can increase payoffs most quickly. 

A new lab with an age of zero is then created, imperfectly inheriting the attributes of its parent lab.  Power, effort, and replication rate independently ``mutate" with probabilities $\mu_w$, $\mu_e$, and $\mu_r$ respectively. If a mutation occurs, the parameter value inherited from the parent lab is increased or decreased by an amount drawn from a Gaussian distribution with a mean of zero and a standard deviation of $\sigma_w$, $\sigma_e$, or $\sigma_r$ for power, effort, or replication rate. If a mutation modifies a parameter's value below or above its prescribed range, it is truncated to the minimum or maximum value. 

It is, of course, unrealistic to assume that all researchers have the same expected ``lifespan." Many researchers disappear from academia quite early in their careers, shortly after receiving their degree or completing a postdoc. Nevertheless, the simplifying assumption of equal expected lifespan is, if anything, a conservative one for our argument. If the factors that lead a researcher to drop out of the field early in his or her career are unrelated to publication, then this is irrelevant to the model---it is simply noise, and incorporated in the stochastic nature of the death algorithm. On the other hand, if the ability to progress in one's career is directly influenced by publications, then our model is, if anything, a muted demonstration of the strength of selection on publication quantity. 

After death and reproduction, the published literature is truncated to a manageable size. Because few fields have more than a million relevant publications (assessed through searches for broad key words in Google Scholar---most fields probably have far fewer relevant papers), because most replications target recent work (e.g., 90\% of replications in psychology target work less than 10 years old \citep{pashler_replicability_2012}), and because decreased data availability for older papers makes replication more difficult \citep{vines2014availability}, we restrict the size of the literature available for replication to the one million most recently published results. This assumption was made in part to keep the computational load at manageable levels and to allow for long evolutionary trajectories. At the end of every evolutionary stage, the oldest papers in the published literature are removed until the total number is less than or equal to one million. 

\begin{table}
\caption{Global model parameters.}
\begin{center}
\begin{tabular}{|c|l|c|}
\hline
{\bf Parameter} 	&	{\bf Definition}	&	{\bf Values tested}\\
\hline
$N$ 			&	Number of labs			& 100	\\
$b$ 			&	Base rate of true hypotheses			& 0.1	\\
$r_0$ 			&	Initial replication rate for all labs			& \{0, 0.01, 0.2, 0.5\}	\\
$e_0$ 			&	Initial effort for all labs			& 75 	\\
$w_0$ 			&	Initial power for all labs			& 0.8 	\\
$\eta$ 			&	Influence of effort on productivity			& 0.2 	\\
$c_{R+}$ 			&	Probability of publishing positive replication			& 1 	\\
$c_{R-}$  			&	Probability of publishing negative replication			& 1 	\\
$V_N$ 			&	Payoff for publishing novel result			& 1 	\\
$V_{R+}$ 			&	Payoff for publishing positive replication			& 0.5	\\
$V_{R-}$  			&	Payoff for publishing negative replication			& 0.5	\\
$V_{O+}$ 			&	Payoff for having novel result replicated			& 0.1 	\\
$V_{O-}$  			&	Payoff for having novel result fail to replicate			& $-100$	\\
$d$ 			&	Number of labs sampled for death and birth evens			& 10	\\
$\mu_r$ 			&	Probability of $r$ mutation			& \{0, 0.01\}	\\
$\mu_e$			&	Probability of $e$ mutation			& \{0, 0.01\}	\\
$\mu_w$			&	Probability of $w$ mutation			& \{0, 0.01\} 	\\
$\sigma_r$ 			&	Standard deviation of $r$ mutation magnitude			& 0.01	\\
$\sigma_e$ 		&	Standard deviation of $e$ mutation magnitude			& 1	\\
$\sigma_w$ 		&	Standard deviation of $w$ mutation magnitude			& 0.01	\\
\hline
\end{tabular}
\end{center}
\label{table_parameters}
\end{table}

\subsection{Simulation runs}
Table \ref{table_parameters} describes all the model parameters and their default and range values. The data are averaged from 50 runs for each set of parameter values tested.  Each simulation was run for one million time steps, with data collected every 2000 time steps. In addition to the evolvable traits of individual research groups, we also collected data on the false discovery rate (FDR) for all positive results published each time step. Our goal here is illustrative rather than exploratory, and so our analysis focuses on a few illuminative examples. For the interested reader, we provide the Java code for the full model as a supplement.

\section{Simulation Results}

While our model is thankfully much simpler than any real scientific community, it is still quite complex. So we introduce bits of the dynamics in several sections, so that readers can learn from the isolated forces.

\subsection{The natural selection of bad science.} 
\begin{figure}[tp]
\begin{center}
	\hspace{-0.5in}\includegraphics[width=4in]{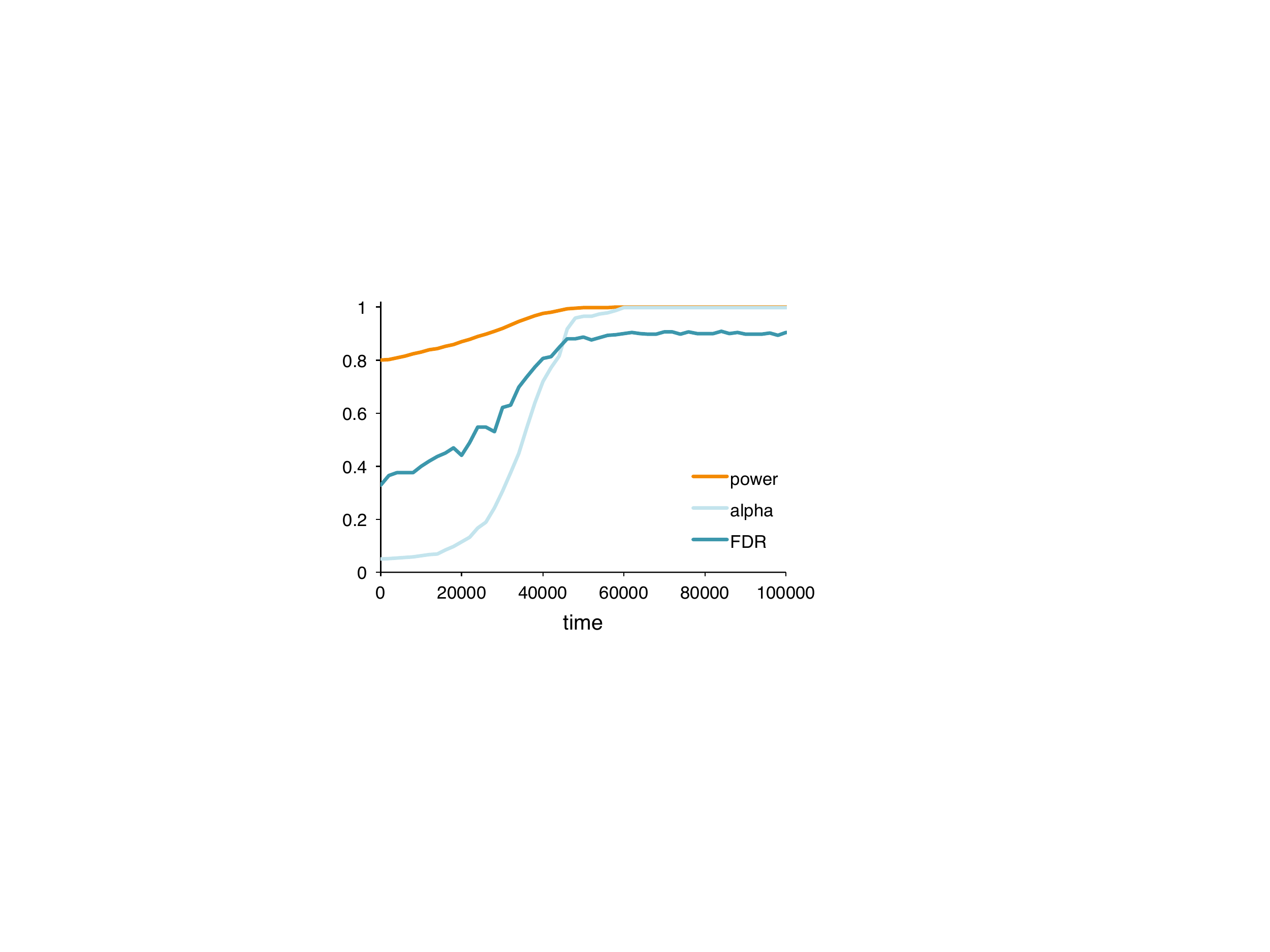}
\caption{Power evolves. The evolution of mean power ($W$), false positive rate ($\alpha$), and  false discovery rate (FDR). }
\label{fig_dyn_power}
\end{center}
\end{figure}

First, ignore replication and focus instead on the production of novel findings ($r_0 = \mu_r = 0$). Consider the evolution of power in the face of constant effort ($\mu_e = 0$, $\mu_w = 0.01$).  As power increases so too does the rate of false positives. Higher power therefore increases the total number of positive results a lab obtains, and hence the number of papers communicated. As such, both $W$ and $\alpha$ go to unity in our simulations, such that {\em all} results are positive and thereby publishable (Figure \ref{fig_dyn_power}).

It is easy to prove that unimpeded power (and false positive rate) will increase to unity. Fitness is directly tied to the number of publications a lab produces, so anything that increases number of publications also increases fitness. The probability that a novel hypothesis is true is the base rate, $b$. For a lab with power $W$ and effort $e$, the probability that a test of a novel hypothesis will lead to a positive (and therefore publishable) result is therefore 
\begin{align}
\Pr(+) 	&= \Pr(+|T) \Pr(T) + \Pr(+|F) \Pr(F) \nonumber \\
		&= bW + (1-b)\frac{W}{1 + (1 - W) e}.
\end{align}
By differentiation, it can be shown that this probability is strictly increasing as a function of $W$.
This implies that if selection favors ever higher discovery rates, power will continue to increase to the point at which it is matched by a countervailing force, that is, by whatever factors limit the extent to which power and false positive rate can change. 

This first case is unrealistic. Research groups would never be able to get away with methods for which {\em all} hypotheses are supported, not to mention that increasing power without bound is not pragmatically feasible. However, one {\em can} imagine institutions to keep power relatively high, insofar as at least some aspects of power are directly measurable. At least some aspects of experimental power, such as statistical power, are directly measurable. False positives, on the other hand, are notoriously difficult  for peer reviewers to assess \citep{macarthur2012methods, eyre2013assessment}.  

\begin{figure}[tp]
\begin{center}
	\hspace{-0.5in}\includegraphics[width=4in]{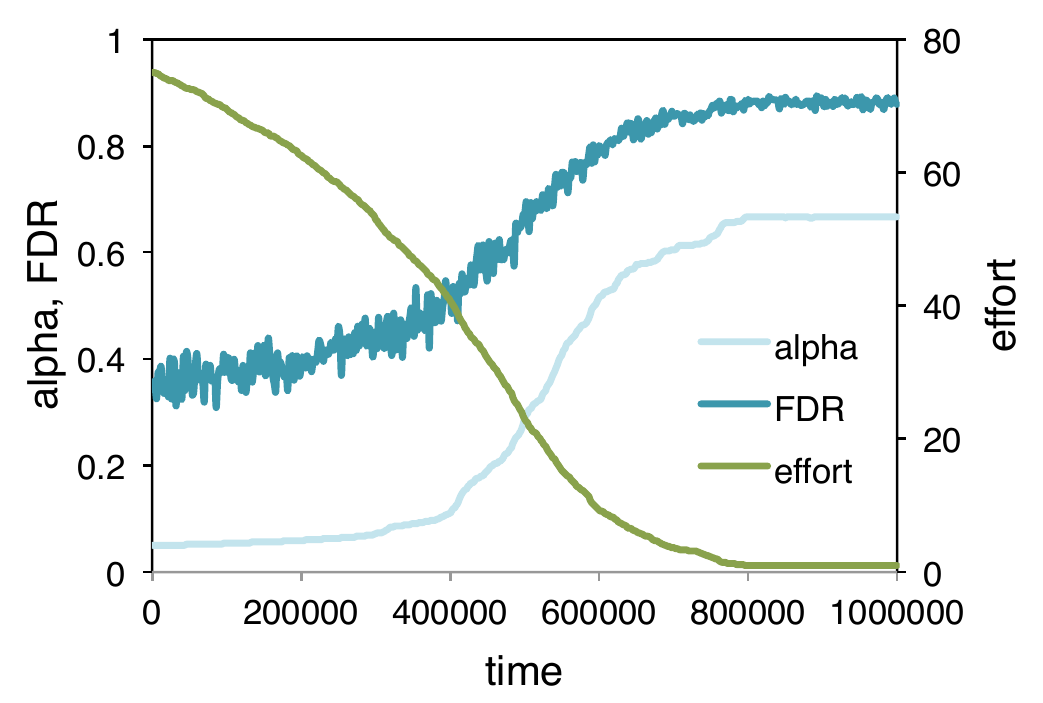}
\caption{Effort evolves. The evolution of low mean effort corresponds to evolution of high false positive and false discovery rates.}
\label{fig_dyn_effort}
\end{center}
\end{figure}

If power is measurable and kept high through institutional enforcement, publication rates can still be increased by reducing the effort needed to avoid false positives. We ran simulations in which power was held constant but in which effort could evolve ($\mu_w = 0$, $\mu_e = 0.01$). Here selection favored labs who put in less effort toward ensuring quality work, which increased publication rates at the cost of more false discoveries (Figure \ref{fig_dyn_effort}). When the focus is on the production of novel results and negative findings are difficult to publish, institutional incentives for publication quantity select for the continued degradation of scientific practices.  

\subsection{The ineffectuality of replication} 
Novel results are not and should not be the sole focus of scientific research. False discoveries and ambiguous results are inevitable with even the most rigorous methods. The only way to effectively separate true from false hypotheses is through repeated investigation, including both direct and conceptual replication \citep{mcelreath_boffland_2015, pashler_replicability_2012}. Can replication impede the evolution of bad science? If replications are difficult to publish, this is unlikely. On the other hand, consider a scenario in which replication efforts are easy to publish and the failure to replicate a lab's novel result causes substantial losses to the prestige of that lab.  Might the introduction of replication then counteract selection for low-effort methodologies?      

We repeated the previous runs, but this time initialized each lab so that 1\% of all investigations would be replications of (randomly selected) hypotheses from the published literature. This is consistent with empirical estimations of replication rates in psychology \citep{makel_replications_2012}. We then allowed the replication rate to evolve through mutation and selection ($r_0 = 0.01$, $\mu_w = 0$, $\mu_r = \mu_e = 0.01$). Conditions in these runs were highly favorable to replication. We assumed that all replication efforts would be publishable, and worth half as much as a novel result in terms of evolutionary fitness (e.g., in terms of associated prestige). Additionally, having one's original novel result successfully replicated by another lab increased the value of that finding by 10\%, but having one’s original result fail to replicate was catastrophic, carrying a penalty equal to 100 times the value of the initial finding (i.e., $V_{O+} = 0.1$, $V_{O-} = -100$). This last assumption may appear unrealistically harsh, but research indicates that retractions can lead to a substantial decrease in citations to researchers' prior work \citep{lu2013retraction}. In addition, some have suggested that institutions should incentivize reproducibility by providing some sort of ``money-back guarantee" if results fail to replicate \citep{rosenblatt2016incentive}, which could end up being highly punitive to individual researchers. More generally, these assumptions are extremely favorable to the idea that replication might deter poor methods, since false positives carry the highest risk of failing to replicate. 

\begin{figure}[tp]
\begin{center}
	\hspace{-0.5in}\includegraphics[width=4in]{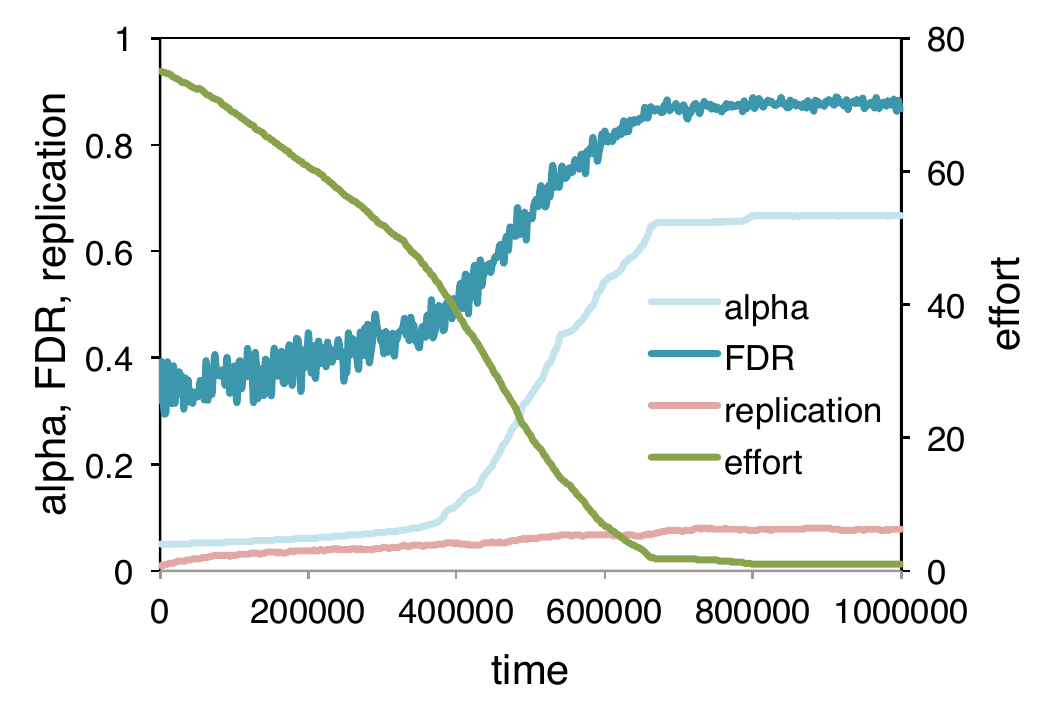}
\caption{The coevolution of effort and replication.}
\label{fig_dyn_replication}
\end{center}
\end{figure}

We found that the mean rate of replication evolved slowly but steadily to around 0.08. Replication was weakly selected for, because although publication of a replication was worth only half as much as publication of a novel result, it was also guaranteed to be published. On the other hand, allowing replication to evolve could not stave off the evolution of low effort, because low effort increased the false positive rate to such high levels that novel hypotheses became more likely than not to yield positive results (Figure \ref{fig_dyn_replication}). As such, increasing one's replication rate became less lucrative than reducing effort and pursuing novel hypotheses. 

\begin{figure}[tp]
\begin{center}
	\hspace{-0.5in}\includegraphics[width=4in]{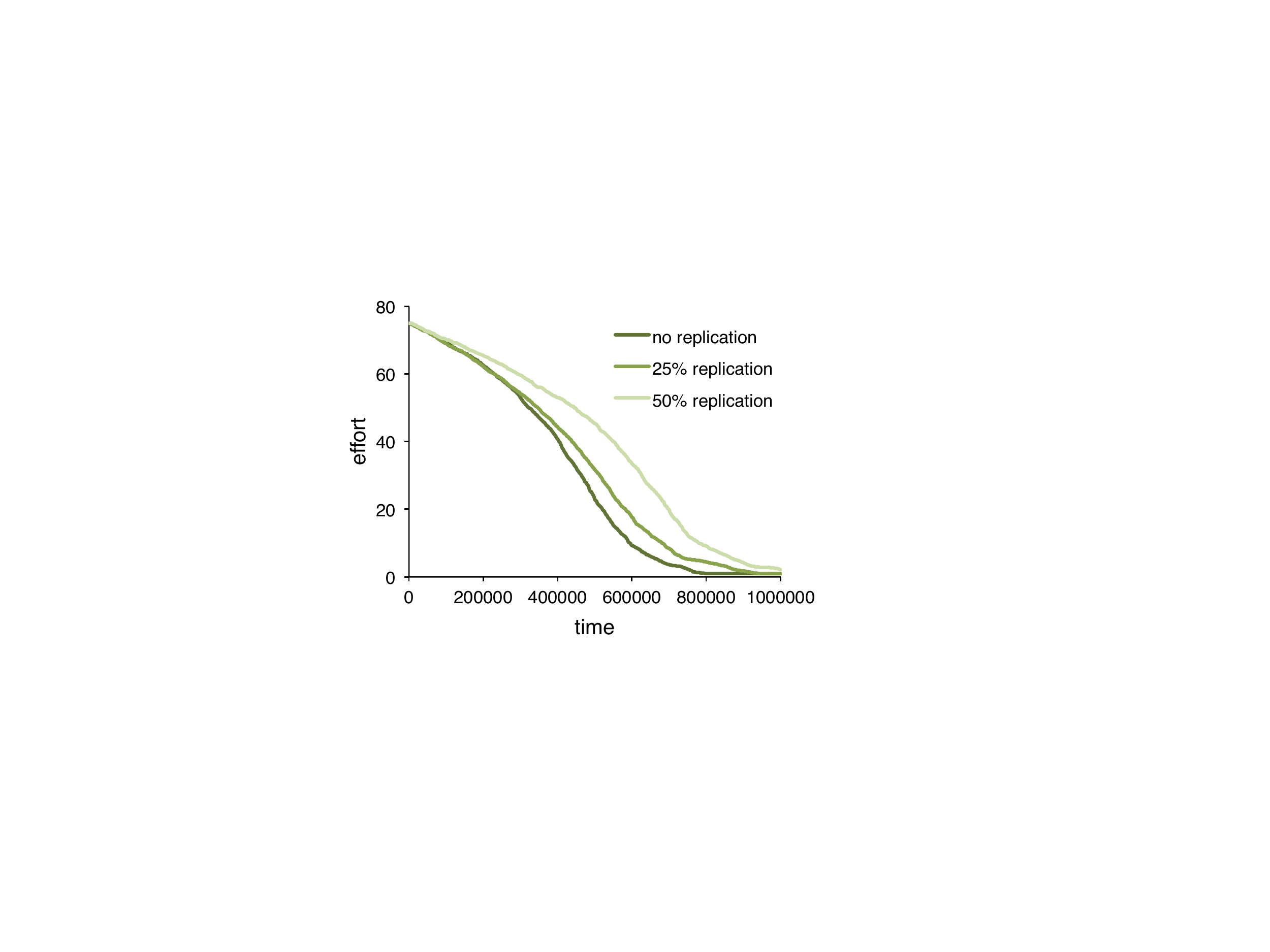}
\caption{The evolution of effort when zero, 25\%, or 50\% of all studies performed are replications.}
\label{fig_dyn_effortVrep}
\end{center}
\end{figure}

Because effort decreased much more rapidly than replication rates increased, we considered the possibility that substantially higher replication rates {\em might} be effective at keeping effort rates high, if only they could be enforced. 
To test this hypothesis, we ran simulations in which the replication rate could not mutate ($\mu_r = 0$) but was initialized to very high levels. High rates of replication did slow the decline of effort, but even extremely high replication rates (as high as 50\%) did not stop effort from eventually bottoming out (Figure \ref{fig_dyn_effortVrep}). We note emphatically that we are not suggesting that highly punitive outcomes for failures to replicate are desirable, since even high-quality research will occasionally fail to replicate. Rather, we are pointing out that even in the presence of such punitive outcomes, institutional incentives for publication quality will still select for increasingly low-quality methods. 

\subsection{Why isn't replication sufficient?} 
Replication is not sufficient to curb the natural selection of bad science because the top performing labs will always be those who are able to cut corners.  Replication allows those labs with poor methods to be penalized, but unless {\em all} published studies are replicated several times (an ideal but implausible scenario), some labs will avoid being caught. In a system such as modern science, with finite career opportunities and high network connectivity, the marginal return for being in the top tier of publications may be orders of magnitude higher than an otherwise respectable publication record \citep{clauset2015systematic, ioannidis2014estimates, frank2012darwin}. 

Within our evolutionary model it was difficult to lay bare the precise relationship between replication, effort, and reproductive success, because all these parameters are entangled. We wanted to understand exactly why the most successful labs appear to be those with low effort even when failure to replicate was highly punitive, making low effort potentially quite costly. To do this, we created a simplified version of our model that omitted any evolution. Power was fixed at 0.8 and all labs were either high effort (H) or low effort (L). High effort labs had effort $e_H = 0.75$, corresponding to a false positive rate of $\alpha = 0.05$ and an investigation rate of $h = 0.625$. Low effort labs had effort $e_L = 0.15$, corresponding to a false positive rate of $\alpha = 0.2$ and an investigation rate of $h = 0.765$. The population was initialized with 50\% high effort labs. Labs conducted research and attempted to communicate their findings as in the original model. We allowed ten time steps without any replication to establish a baseline body of literature, and then ran the simulation for 100 additional time steps, equivalent to the expected lifespan of a lab in the evolutionary model. During this time, each hypothesis investigated was a replication with probability $r$. This simplified model allowed us to examine the distribution of the payoffs (resulting from both the benefits of successful publications and punishment from failed replications) and to directly compare high and low effort labs. 

\begin{figure}[tp]
\begin{center}
	\hspace{-0.0in}\includegraphics[width=5.9in]{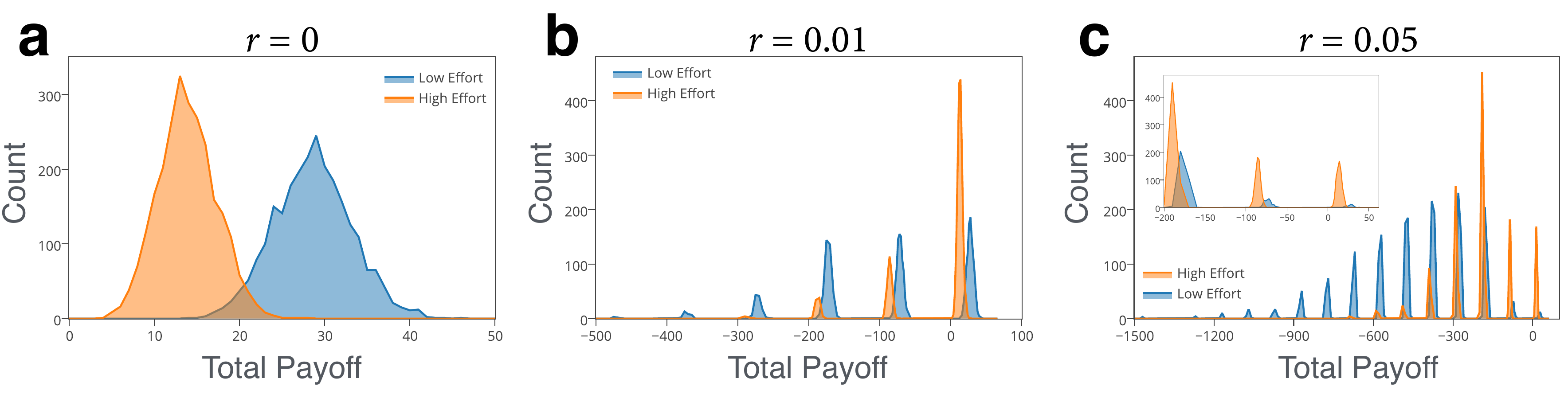}
\caption{Lab payoffs from the non-evolutionary model. Each graph shows count distributions for high and low effort labs' total payoffs after 110 time steps, 100 of which included replication. Total count for each payoff is totaled from 50 runs for each condition. Subfigure (c) includes an inset that displays the same data as the larger graph, but for a narrower range of payoffs.}
\label{fig_extrafitness}
\end{center}
\end{figure}

Figure \ref{fig_extrafitness} shows the distributions of payoffs from three replication rates. Without replication, low effort labs have an unambiguous advantage. As the rate of replication increases, the mean payoff for high effort labs can surpass the mean payoff for low effort labs, as the former are less likely to be punished. However, the laws of probability dictate that {\em some} labs will escape either producing false positives or being caught doing so, and among these the very highest performers will be those who exert low effort. 
When the top labs have disproportionate influence on funding and graduate student success, this type of small advantage can cascade to continuously select for lower effort and increasing numbers of false discoveries, as seen in our evolutionary model.

\section{Discussion}

Incentives drive cultural evolution. In the scientific community, incentives for publication quantity can drive the evolution of poor methodological practices. We have provided some empirical evidence that this occurred, as well as a general model of the process. 
If we want to improve how our scientific culture functions, we must consider not only the individual behaviors we wish to change, but also the social forces that provide affordances and incentives for those behaviors \citep{Campbell:1976, wilson2014bbs}. We are hardly the first to consider a need to alter the incentives for career success in science \citep{nosek_2014, nosek2012scientific, ioannidis2014make, vankov2014, fischer2012academia, brembs2013deep, begley2015incentives, maccoun2015blind, gigerenzer2015surrogate,sills2016success, sarewitz2016pressure}. However, we are the first to illustrate the evolutionary logic of how, in the absence of change, the existing incentives will {\em necessarily} lead to the degradation of scientific practices. 

An incentive structure that rewards publication quantity will, in the absence of countervailing forces, select for methods that produce the greatest number of publishable results. This in turn will lead to the natural selection of poor methods and increasingly high false discovery rates. Although we have focused on false discoveries, there are additional negative repercussions of this kind of incentive structure. Scrupulous research on difficult problems may require years of intense work before yielding coherent, publishable results. If shallower work generating more publications is favored, then researchers interested in pursuing complex questions may find themselves without jobs, perhaps to the detriment of the scientific community more broadly.  

Good science is in some sense a public good, and as such may be characterized by the conflict between cooperation and free riding. We can think of cooperation here as the opportunity to create group-beneficial outcomes (i.e., quality research) at a personal cost (i.e., diminished ``fitness" in terms of academic success). To those familiar with the game theory of cooperative dilemmas, it might therefore appear that continued contributions to the public good---cooperation rather than free riding---could be maintained through the same mechanisms known to promote cooperation more generally, including reciprocity, monitoring, and punishment \citep{rand2013coop}. However, the logic of cooperation requires that the benefit received by cooperators can be measured in the same units as the payoff to free riders: i.e., units of evolutionary fitness. It is possible that coalitions of rigorous scientists working together will generate greater output than less rigorous individuals working in isolation. And indeed, there has been an increase in highly collaborative work in many fields \citep{nabout2015publish, wardil2015coauthor}. Nevertheless, such collaboration may also be a direct response to incentives for publication quantity, as contributing a small amount to many projects generates more publications than does contributing a large amount to few projects. Cooperation in the sense of higher quality research provides a public good in the sense of knowledge, but not in the sense of fitness for the cultural evolution of methodology. Purely bottom-up solutions are therefore unlikely to be sufficient. That said, changing attitudes about the assessment of scientists is vital to making progress, and is a driving motivation for this presentation. 

Institutional change is difficult to accomplish, because it requires coordination on a large scale, which is often costly to early adopters \citep{north1990institutions, aoki2007endogenizing}. Yet such change is needed to insure the integrity of science. It is therefore worth considering the types of institutions we might want. 
It might appear that journals and peer reviewers need only to adopt increasingly strict standards as bars to publication. However, although this may place some limits on the extent to which individuals can game the system, it still affords the opportunity to do so \citep{martin2016impact}, and {\em incentives} to do so will exist as long as success is tied to publication.  

Punishing individuals for failure to replicate their original results is unlikely to be effective at stopping the evolution of bad science. Many important discoveries initially appear unlikely, and designing clear experiments is difficult. Eliminating false positives entirely is likely to be impossible \citep{maccoun2015blind}.  In addition, failed replications may themselves be false negatives. Moreover, replication is difficult or impossible for some fields, such as those involving large clinical samples or historical data. An overemphasis on replication as the savior of all science is biased in favor of certain fields over others. It is therefore inadvisable to be overly harsh on researchers when their results fail to replicate. 

A more drastic suggestion is to alter the fitness landscape entirely by changing the selection pressures: the incentives for success. This is likely to be quite difficult.  Consider that the stakeholders who fund and support science may have incentives that are not always aligned with those of active researchers \citep{ioannidis2014make}. For example, funders may expect ``deliverables" in the form of published papers, which in turn may pressure scientists to conduct research in such a manner to maximize those deliverables, even if incentives for promotion or hiring are changed. Another impediment is the constraint on time and cognitive resources on the part of evaluators. The quality of a researcher is difficult to assess, particularly since there are many ways to be a good scientist, making assessment a high-dimensionality optimization problem. Quantitative metrics such as publication rates and impact factor are used partly because they are simple and provide clear, unambiguous comparisons between researchers. Yet these are precisely the qualities that allow such metrics to be exploited. In reality, the true fitness landscape of scientific career success is multidimensional. Although some publications are probably necessary, there are other routes to success beside the accrual of a lengthy CV. We should support institutions that facilitate these routes, particularly when they encourage high quality research, and resist the temptation or social pressure to paper count. 

Our model treats each publication of a novel result as equivalent, but of course they are not. Instead, the most prestigious journals receive the most attention and are the most heavily cited \citep{ioannidis2006concentration}. Anecdotally, we have heard many academics --- ranging from graduate students to full professors --- discussing job candidates largely in terms of the prestigious journals they either published in or failed to publish in. 
Consideration of journal prestige would change our model somewhat, but the implications would be similar, as the existence of such journals creates pressure to publish in those journals at all costs. The major change to our model would be additional selection for highly surprising results, which are more likely to be false.  Investigations have found that the statistical power of papers published in prestigious (high impact factor) journals is no different from those with lower impact factors \citep{brembs2013deep}, while the rate of retractions for journals is positively correlated with impact factor \citep{fang2011retracted}. Although this is likely to be at least partly due to the increased attention paid to those papers, it is well known that high impact journals often reject competent papers deemed insufficiently novel.

Our model presents a somewhat pessimistic picture of the scientific community. Let us be clear: many scientists are well aware of the difficulties surrounding evaluation, and many hiring and grant committees take efforts to evaluate researchers on the quality of their work rather than the quantity of their output or where it is published. Moreover, we believe that many if not most scientists are truly driven to discover truth about the world. It would be overly cynical to believe that scientists are driven only by extrinsic incentives, fashioning researchers to conform to the expectations of {\em Homo economicus}. However, a key feature of our evolutionary model is that it assumes no ill intent on the part of the actors, and does not assume that anyone actively alters their methods in response to incentives. Rather, selection pressures {\em at the institutional level} favored those research groups that, for whatever reason, used methods that generated a higher volume of published work. Any active social learning, such as success-biased copying, will serve only to accelerate the pace of evolution for such methods.

Despite incentives for productivity, many scientists employ rigorous methods and learn new things about the world all the time that are validated by replication or by being effectively put into practice. 
In other words, there is still plenty of good science out there. 
One reason is that 
publication volume is rarely the only determinant of the success or failure of a scientist's career. Other important factors include the importance of one's research topic, the quality of one's work, and the esteem of one's peers. The weight of each factor varies among disciplines, and in some fields such factors may work positively to promote 
behaviors leading to high-quality research, particularly when selection for those behaviors is enculturated into institutions or disciplinary norms. 
In such cases, this may be sufficient to counteract the negative effects of incentives for publication volume, and so maintain high levels of research quality. If, on the other hand, success is largely determined by publication output or related quantitative metrics, then those who care about quality research should be on high alert. In which direction the scale tips in one's own field is a critical question for anyone interested in the future of science. 

Whenever quantitative metrics are used as proxies to evaluate and reward scientists, those metrics become open to exploitation if it is easier to do so than to directly improve the quality of research.  Institutional guidelines for evaluation at least partly determine how researchers devote their energies, and thereby shape the kind of science that gets done. 
A real solution is likely to be patchwork, in part because accurately rewarding quality is difficult\footnote{Not to mention that any measure of quality is partly dependent on ever-changing disciplinary, institutional, and departmental needs.}. Real merit takes time to manifest, and scrutinizing the quality of another's work takes time from already busy schedules. Competition for jobs and funding is stiff, and reviewers require some means to assess researchers. Moreover, individuals differ on their criteria for excellence.  Boiling an individual's output to simple, objective metrics, such as number of publications or journal impacts, entails considerable savings in terms of time, energy, and ambiguity. Unfortunately, the long-term costs of using simple quantitative metrics to assess researcher merit are likely to be quite great. If we are serious about ensuring that our science is both meaningful and reproducible, we must ensure that our institutions incentivize that kind of science.

\section*{Acknowledgments}
For critical feedback on earlier drafts of this manuscript, we thank Clark Barrett, William Baum, Monique Borgerhoff Mulder, John Bunce, Emily Newton, Joel Steele, Peter Trimmer, and Bruce Winterhalder.

\bibliographystyle{newapa} 
\bibliography{boffincentives}

\end{document}